\documentclass{article}
\newcommand{\ket}[1]{| #1 \rangle}
\newcommand{\bra}[1]{\langle #1 |}

\newcommand{\Op}[1]{\hat{#1}}
\newcommand{\Hil}{\mathcal{H}}

\newcommand\eq[1]{Eq.~(\ref{#1})}

\usepackage{amsmath}
\title{\bf Nondeterministic Recursion with Quantum Units}
\date{}
\author{\em Alexander Yu.\ Vlasov}
\begin{document}
\maketitle
\begin{abstract}
 In this paper are discussed some formal properties of quantum devices necessary
 for implementation of nondeterministic Turing machine.
\end{abstract}
\sloppy

\section{Introduction}

 Classical {\em nondeterministic Turing machine} has possibility to produce few branches
of computing process and so resolve some exponential problem in linear time. 
Despite of possibility of modeling such device using conventional software,
any real speedup may be implemented only if new instances of ``hardware'' could be 
created with an appropriate rate in comparison with speed of calculations.

Using some self-reproducing nano-devices with sizes comparable with molecular one 
could be reasonable for such a purposes, and laws of quantum mechanics may be 
relevant for such a case. So present paper is devoted to rather old theme of 
``quantum self-reproducing or cloning'' \cite{QSelf,QClon} from point of view 
of some recent achievements.

For simplicity the Schr\"odinger picture is used in present paper and it is
shown, that respect of quantum laws does not produce some crucial limitations
on construction of self-reproducing nano-devices with quite universal properties. 

\subsection*{Preliminaries} 

Two fundamental works: Wigner (1961) \cite{QSelf}, Wootters and Zurek (1982) 
\cite{QClon} produce a good basis for the theme. In second paper \cite{QClon} 
is considered only example with two-dimensional space of photon polarizations, 
but method used in the work may be reformulated to Hilbert space with dimension 
$n$ and in such a case a ``positive part'' of no-cloning theorem can be formulated as: 
{\em it is possible to clone no more than $n$ different orthogonal states between 
$n$-dimensional continuum of all possible states.}

So we can see, that even in best case only set of states with measure zero
could be cloned, and only for some special choice of interaction. It is
in good agreement with initial idea of Wigner \cite{QSelf}, that probability 
of self-reproducing unit for general setup is vanishing.

Fortunately, theme of present paper does not include deep questions about
probability of appearance, evolution, stability of quantum automata, etc. 
Such questions are usualy not considered in computer science.
It is simply suggested, that there is an automata with certain structure 
and question is construction of a branching process with new and new instances 
of initial sample. 

It is also suggested, that quantum effect may not be neglected in our design
due to size, structure of automata or other reason. 
 
\section{Cloning of orthogonal states}
 
No-cloning theorem save possibility to clone $n$ orthogonal states, but Wigner
consideration show, that evolution with such property is not very common, 
but anyway is possible. Let us construct example of such process.
 
Let us consider some basis in $n$-dimensional Hilbert space $\Hil$. We will use
Dirac notation $\ket{k}$, $k = 0,\ldots,n-1$ for elements of the basis
and $\ket{k}\bra{k}$ for projector on element $\ket{k}$. We want to describe
unitary operator, that clones only $n$ elements --- the given basis. It is
known, that no-cloning theorem does not forbid it.

Let us consider unitary operator of cyclic permutations
of all basis vectors together with all powers of it
\begin{equation}
\Op U \ket{k} = \ket{k+1 \bmod n},
\quad
\Op U^l \ket{k} = \ket{k+l \bmod n}.
\end{equation}
    
It is simple to check, that if there is composite system described as tensor
product of two Hilbert spaces $\Hil \otimes \Hil$, then operator described as
\begin{equation}
 \Op C = \sum_l \bigl(\ket{l}\bra{l} \otimes \Op U^l\bigr)
\label{CondClon}
\end{equation}
has property
\begin{equation}
 \Op C : \ket{k}\otimes\ket{0} \longrightarrow \ket{k}\otimes\ket{k},
 \quad k = 0,\ldots,n-1.
\label{nClon}
\end{equation}

Here $\ket{0}$ is fixed state of ``environment'' of the quantum automata before
``cloning.'' Following to idea\footnote{In simpler case then state of apparatus 
are same for any initial state of photon.} of \cite{QClon} it is possible to show,
that due to linearity of quantum mechanics \eq{nClon} define action on arbitrary 
superposition $\ket{\psi}=\sum_k \psi_k \ket{k}$ and such action differs 
from {\em nonlinear} expression, $\ket{\psi}\ket{0} \not\to \ket{\psi}\ket{\psi}$. 
On the other hand, nonlinear and linear function may coincide in fixed number of points,
and we just show existence of $n$ such points.

\section{Conditional quantum dynamics}

The construction \eq{CondClon} used above
is particular example of {\em conditional quantum dynamics} \cite{Joz95},%
\footnote{
It should be mentioned here, that despite of some results and methods used
in present text often associated with {\em quantum information science},
current presentation does not suggest necessity of close familiarity with
this area.}
there instead of power of $\Op U$ are used set of arbitrary unitary operators 
$\Op U_l$ \cite{Joz95} 
\begin{equation}
 \Op D = \sum_l \bigl(\ket{l}\bra{l} \otimes \Op U_l\bigr),
\label{CondDyn}
\end{equation}
so instead of \eq{nClon} it may be written
\begin{equation}
 \Op D : \ket{k}\otimes\ket{0} \longrightarrow \ket{k}\otimes\bigl(U_k\ket{0}\bigr).
\label{prArr}
\end{equation}
Dimensions of Hilbert spaces in conditional quantum dynamics are not
necessary coincide, $\ket{k}\otimes\ket{0} \in \Hil_1 \otimes \Hil_2$,
$\dim \Hil_1 = n$, $\dim \Hil_2 = m$.

Last expression \eq{prArr} is also known in more general form due to application
to theory of {\em programmable} quantum gates \cite{NC97}, but here 
it is useful for consideration of self-reproducing quantum
automata. Really, it was already mentioned, that $\Op U_k$ may be arbitrary 
$n$ unitary operators and so $\Op U_k\ket{0}$ may be any
$n$ quantum states, {\em not necessary orthogonal}. 

It is clear from such construction, that number of different operators 
$\Op U_k$ coincides with dimension $n$ of first Hilbert spaces and it is known 
\cite{NC97}, that same limitation valid for any operator reproducing
programmable quantum dynamics with pure states.

\section{Universal approximation}

Dynamics \eq{CondDyn} is particularly useful, if set of operators $\Op U_k$
is universal \cite{Deu85}, i.e. any unitary operator may be expressed or
approximated with necessary precision as products of operators $\Op U_k$.
It was already discussed, that number of such operators is limited by
dimension of Hilbert space $n$ and the number is finite in our model,
but more general model of programmable quantum dynamics let us partially
bypass such limitation.

It is enough to consider, that we have some long tape with different states
and two unitary operators: first one may shift the tape on one cell, and second
one is operator $\Op D$ \eq{CondDyn} applied to current cell. It is clear, that 
consequent application of such process is equivalent to application of
series operator $\Op U_k$ to state of second system \cite{Vla123}
\begin{equation}
\Op D^s \bigl(\ket{k_s \ldots k_1}\otimes\ket{0}\bigr) = 
\ket{k_s \ldots k_1}\otimes\bigl(\Op U_s \cdots \Op U_1\ket{0}\bigr),
\label{qTape}
\end{equation}
here is considered cyclic shift of tape with length $s$ and so after 
$s$ steps it is returned to initial state, but second system suffers evolution
``encoded'' in first system. It should be mentioned
also, that all states of tape are orthogonal states, it is $n^s$ possible tensor 
products of initial basis for first system.

It was already mentioned \cite{NC97} that it is not possible to express any operator
by such a way, but there are products of operators, that may approximate it
with arbitrary precision \cite{Deu85} if sequence is long enough. 
It is called sometime {\em universality in approximate sense}.
So using quite long tape, we may encode arbitrary state of second system.

\section{Self-reproducing units}

Using methods above, self-reproducing quantum units could have
following structure. First, there is subsystem with description of structure
of the automata $\ket{T}$ and all necessary operations, like ``tape'' \eq{qTape}. 
All used states of the subsystem are orthogonal\footnote{At least at 
moment of time preceding creating of copy of automata.} and so may be precisely 
copied. In addition the automata must have possibility to perform two
specific operations: first one is 
$\Op C$ \eq{CondClon} for producing of two copies of tape and second one
is $\Op D$ \eq{CondDyn} for preparation of arbitrary quantum state with
necessary precision. 

Process of creation of derived structure may be considered by following steps.
At first, it is ``replication'' of state $\ket{T}$, using operator \eq{CondClon} with
each segment initialized by $\ket{0}$ and shifts of the tape, segment by segment.
Such operation is not prohibited by no-cloning theorem, because all different
states of tape are orthogonal.

Of course, most subtle question is how to supply derived structure with operators
$\Op{C}$ and $\Op{D}$ necessary for further functioning of the automata.

A naive idea is that in real design an operator is not some ``external''
object, but result of interactions of different parts of automata, i.e. also may
be encoded in states of different structures, more formally, instead of
action of some operator $\ket{\psi'}= \Op G \ket{\psi}$, it is considered
process
\begin{equation}
 \Op{\mathcal{S}} : \ket{\Psi_G}\otimes \ket{\psi} 
 \longrightarrow  \ket{\Psi_G}\otimes(G \ket{\psi}),
\end{equation}
there $\ket{\Psi_G}$ is state of all parts of given automata necessary for
implementation of operator $\Op G$, and $\Op{\mathcal{S}}$ is fixed formal scattering 
operator taking into account all dynamical laws\footnote{Operator describing
``The Laws of Nature.''}.

From such point of view for construction of operators $\Op C$ and $\Op D$ it
is only necessary to have possibility to create some states $\Psi_C$ and 
$\Psi_D$, but it is known states and so always may be constructed with necessary 
precision using ``universal translator'' \eq{qTape} with $\Op D$ of parent automata.
It is only necessary to have algorithms of construction of $\Psi_C$ and $\Psi_D$ encoded 
by some parts of tape $\ket{T}$.

\medskip

Finally, process of creation of copy for given quantum automata may be described
as following.

\begin{enumerate}
\item Replication of tape $\ket{T}$. The tape contains sequences encoding 
 $\ket{\Psi_C}$, $\ket{\Psi_D}$, etc.
\item The tape is translated to states $\ket{\Psi_C}$, $\ket{\Psi_D}$, etc., 
 using operator $\Op D$ of parent automata.
\end{enumerate}

More generally, any state of such quantum automata may be
described as $\ket{T} \otimes \ket{\Phi_T}$, there $\ket{\Phi_T}$ is ``translation''
of tape $\ket{T}$ using \eq{qTape}. All possible states of tape $\ket{T}$ are
orthogonal, but it is not necessary so for $\ket{\Phi_T}$. Anyway all possible states of
automata are orthogonal due to standard property of scalar and tensor products. 
It explains, why propagation of such automata is not contradict to no-cloning theorem.


\begin{thebibliography}{9}
\bibitem{QSelf} E. P. Wigner,
``The Probability of the Existence of a Self-Reproducing Unit,'' in
{\em The Logic of Personal Knowledge. Essays in Honor of Michael Polanyi},
(Routledge and Kegan Paul, London, 1961);
reprinted in E. P. Wigner,
{\em Symmetries and Reflections}, (Indiana
University Press, Bloomington, Indiana, 1967).
\bibitem{QClon} W. K. Wootters and W. H. Zurek,
``A single quantum cannot be cloned,''
{\em Nature} {\bf 299}, 802--803 (1982).
\bibitem{Joz95} A. Barenco, D. Deutsch, A. K. Ekert, and R. Jozsa,
 ``Conditional quantum dynamics and logic gates,''
 {\em Phys. Rev. Lett.} {\bf 74}, 4083--4086 (1995).
\bibitem{NC97} M. A. Nielsen and I. L Chuang, ``Programmable quantum
 gate arrays,'' {\em Phys. Rev. Lett.} {\bf 79} 321--324, (1997).
\bibitem{Deu85} D. Deutsch, ``Quantum theory, the Church-Turing principle
and the universal quantum computer,''
{\em Proc. R. Soc. London A} {\bf 400}, 97--117 (1985).
\bibitem{Vla123} 
 A. Yu. Vlasov, ``Universal quantum processors with
 arbitrary radix $n \ge 2$,'' {\em Proc. Int. Conf. Q. Inf. (ICQI)}, 
(Rochester 2001), {\em Preprint quant-ph/0103127};
\newblock ---, ``Universal hybrid quantum  processors,'' 
{\em Part. Nucl., Lett.} {\bf 116}, 60--65 (2003),
{\em Preprint quant-ph/0205074};
\newblock ---, ``Quantum processors and controllers,''
 {\em Proc. Int. Conf. Phys. Control}, 861--866 (St-Petersburg 2003),
 {\em Preprint quant-ph/0301147}.
\end{thebibliography}
\end{document}